\documentclass[11pt]{article}
\usepackage[sc]{mathpazo}
\usepackage{fullpage}
\usepackage[authoryear,sectionbib,sort]{natbib}
\usepackage{graphicx}
\usepackage{adjustbox}
\usepackage[boxruled,linesnumbered]{algorithm2e}
\usepackage{multirow}
\usepackage{amsmath}
\usepackage[utf8]{inputenc}
\usepackage{lineno}
\usepackage{titlesec}
\usepackage{graphicx}
\usepackage{subcaption}
\usepackage{booktabs,multirow}
\usepackage{here}
\usepackage{url}

\titleformat{\section}[block]{\Large\bfseries\filcenter}{\thesection}{1em}{}
\titleformat{\subsection}[block]{\Large\itshape\filcenter}{\thesubsection}{1em}{}
\titleformat{\subsubsection}[block]{\large\itshape}{\thesubsubsection}{1em}{}
\titleformat{\paragraph}[runin]{\itshape}{\theparagraph}{1em}{}[. ]

\title{Combining supervised and unsupervised learning methods to predict financial market movements}

\author{Gabriel R. Palma$^{1, 2,\ast}$ \and 
Mariusz Skoczeń$^{4}$ \and
Phil Maguire$^{3}$}

\date{}

\begin{document}

\maketitle

\noindent{} 1. Hamilton Institute, Maynooth University, Maynooth, Ireland;

\noindent{} 2. Department of Mathematics and Statistics, Maynooth University, Maynooth, Ireland;

\noindent{} 3. Department of Computer Science, Maynooth University, Maynooth, Ireland;

\noindent{} 4. DLT Capital, Maynooth, Ireland

\noindent{} $\ast$ Corresponding author; e-mail: gabriel.palma.2022@mumail.ie

\bigskip


\bigskip


\bigskip

\textit{Manuscript type}: Research paper. 

\bigskip

\noindent{\footnotesize Prepared using the suggested \LaTeX{} template for \textit{Am.\ Nat.}}

\newpage{}

\section*{Abstract}
The decisions traders make to buy or sell an asset depend on various analyses, with expertise required to identify patterns that can be exploited for profit. In this paper we identify novel features extracted from emergent and well-established financial markets using linear models and Gaussian Mixture Models (GMM) with the aim of finding profitable opportunities. We used approximately six months of data consisting of minute candles from the Bitcoin, Pepecoin, and Nasdaq markets to derive and compare the proposed novel features with commonly used ones. These features were extracted based on the previous $59$ minutes for each market and used to identify predictions for the hour ahead. We explored the performance of various machine learning strategies, such as Random Forests (RF) and K-Nearest Neighbours (KNN) to classify market movements. A naive random approach to selecting trading decisions was used as a benchmark, with outcomes assumed to be equally likely. We used a temporal cross-validation approach using test sets of $40\%$, $30\%$ and $20\%$ of total hours to evaluate the learning algorithms' performances. Our results showed that filtering the time series facilitates algorithms' generalisation. The GMM filtering approach revealed that the KNN and RF algorithms produced higher average returns than the random algorithm.

\newpage{}

\section{Introduction}

Financial markets are complex systems driven by the collective behaviour and decision-making of market participants \citep{prasad2021,buczynski2021, granha2023three}. The decisions these participants make to buy or sell an asset depend on various analyses, with expertise required to identify patterns that can be exploited for profit \citep{singh2024time, singh2024ensemble}. These patterns can be identified through multiple means such as technical indicators, candlestick charts, or time series statistics \citep{singh2024ensemble, parente2024}. Professional traders hunt for exploitable patterns across all liquid markets including well-established exchanges such as the Nasdaq stock market, the second largest exchange by capitalisation in the USA, as well as emergent markets. One such emergent market is the cryptocurrency domain, which boasts a diverse set of digital coins that can be highly correlated, from meme coins such as Pepe the frog to the classic Bitcoin \citep{sebastiao2021forecasting, fang2022cryptocurrency, cheng2023general}. 

In recent years, machine learning techniques have shown promising results in extracting insights from financial data to forecast market movements and inform trading strategies \citep{GERLEIN2016, dash2016hybrid, paiva2019decision, hasan2020, JAQUART202145, lin2021improving, prasad2021, buczynski2021, SOKOLOVSKY2023, parente2024, BOUTESKA2024, kwon2024hybrid, zou2024novel, avramelou2024deep}. These techniques have been applied in several markets from the various cryptocurrencies \citep{sebastiao2021forecasting, fang2022cryptocurrency, JAQUART202145, parente2024, BOUTESKA2024} to the Istanbul Stock Exchange (BIST100) national index \citep{hasan2020} and other stocks markets \citep{dash2016hybrid, prasad2021, lin2021improving, zou2024novel}. These applications have highlighted the challenges of predicting the future direction of financial markets \citep{buczynski2021}. 


The difficulty of creating price forecasting systems capable of guiding human trading decisions has motivated studies in feature engineering and the design of associated trading strategies \citep{buczynski2021, BOUTESKA2024}. Other authors have examined the use of clustering analysis for understanding financial trends and relationships among financial assets based on unsupervised learning methods, such as Gaussian Mixture Models (GMM) \citep{de2014dynamic, bai2014clustering}, portfolio optimisation \citep{paiva2019decision, cheng2023general}, as well as machine learning algorithms, such as deep neural networks, support vector machine, cascaded lstm networks, ensemble algorithms and others \citep{GERLEIN2016, zou2024novel, singh2024ensemble, parente2024}.

The results of these studies have indicated a need for more sophisticated feature engineering approaches to capture subtle market dynamics. In this paper, we propose a new set of features designed based on the slope and intercept of a linear model fitted to price peaks. The curvature of each peak is used as an explanatory variable based on the close price of the Nasdaq, Pepecoin and Bitcoin markets to illustrate the use of these features following the method proposed by \cite{palma2023}. Moreover, inspired by previous findings of the use of Gaussian Mixture Models (GMMs) as a pre-processing technique applied to time series \citep{eirola2013gaussian, fan2022preprocessing, fan2024using}, we use GMMs to filter market data, checking for scenarios where machine learning algorithms might produce profit. The target is to decide between buying, selling and holding each hour, with approximately $6$ months of data for the three markets. 


We compared the performance of commonly used machine learning algorithms to make trading decisions using a traditional set of features used in previous work \citep{parente2024}. A baseline naive algorithm which made decisions randomly, each with the same probability, was used to aid the comparison of the various algorithms. The proposed approach serves as a basis for further development in feature engineering and pre-processing techniques applied to financial markets.



The rest of this paper is organized as follows. In Section 2, we detail the data acquisition, labelling algorithm, feature engineering and the approach used to cluster the time series based on the set of proposed features. In Section 3, we present and discuss the results obtained. Finally, in Section 4 we present the main conclusions of the study and directions for future work.

\section{Methods}

\subsection*{Data acquisition and labelling algorithm}
We obtained approximately six months of data from Bitcoin and Pepecoin markets, with minute candles describing the open, close, high and low, using the OKX API. Also, we used the backtestmarket API to collect minute candles for the Nasdaq market in the same period. Let $y_{\text{close}}(t)$ be an observation of the close price at minute $t$ for a time series $\mathbf{Y}_m$ for a market $m$. We first obtained the percentage change by computing $ \omega_m(t) = \log{\frac{y_{\text{close}}(t+1)}{y_{\text{close}}(t)}}$ for every market $m$, thus producing time series of percentage changes $\mathbf{\Omega}_1$, $\mathbf{\Omega}_2$ , $\mathbf{\Omega}_3$ for Bitcoin, Pepecoin and Nasdaq respectively. 

\begin{figure}
    \centering
    \includegraphics[width=1\linewidth]{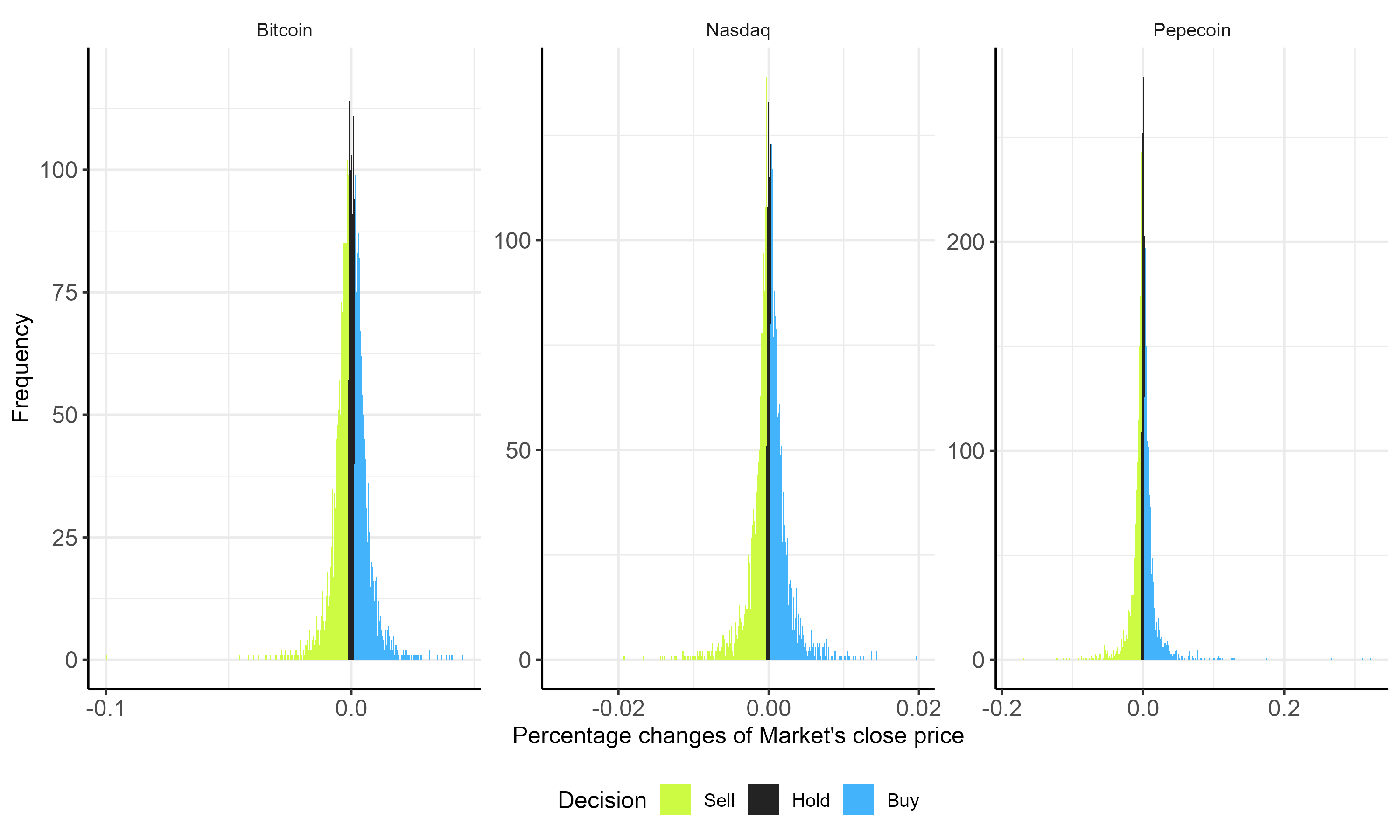}
    \caption{Illustration of the trading decisions obtained from a simple symmetric threshold algorithm based on the $4\%$ quantile of the $\mathbf{\Omega}_m$.}
    \label{DecisionsLabel}
\end{figure}

We defined trading decisions for all markets by employing a simple symmetric threshold algorithm based on the $4\%$ quantile of the $\mathbf{\Omega}_m$. We selected the specified quantile to produce a balanced dataset of market movements. The selected quantile yields a negative $\mathbf{\Omega}_m$ that is defined as the lower threshold, whereas the absolute value is defined as the upper threshold. Thus, if $\omega_m(t)$ is smaller than the lower threshold, we categorize a selling decision for the minute $t$ of the time series $m$. We categorize a buying decision if $\omega_m(t)$ is higher upper threshold. Otherwise, we categorize a holding decision for the minute $t$ of the time series $m$. Figure~\ref{DecisionsLabel} shows an illustration of the results of the threshold algorithm for the studied markets. By following these criteria we obtained a time series of market movements $\mathbf{D}_m$ for Bitcoin, Pepecoin and Nasdaq markets. 

\subsection*{Feature engineering}
For each market, the time series was organised so that for every $60$ observations (minutes), we extracted features using the first $59$ minutes to predict the $60$th observation. We extracted the buy proportion, sell proportion, close price, linear model intercept, linear model slope, peaks average curvature, peaks average magnitude, and estimated the percentage change following the previous description for each market $m$. The linear model parameters were estimated using an adaptation of the features extraction method applied for the EEG time series proposed by \cite{palma2023}. We first collected the peaks of the close price using every $59$ minute interval and collected the estimated curvatures of each peak calculated by taking the second-order difference of $y_{\text{close}}(k-1)$, $y_{\text{close}}(k)$, and $y_{\text{close}}(k+1)$, where $k$ is the time where a peak occurred. Therefore, we fitted a linear model to the peak heights using the curvature of each peak as an explanatory variable and obtained the overall peak average magnitude. Finally, we collected the estimated percentage for each $59$ minutes interval by computing the average of the last $6$ values, $\bar{y}_{\text{close}}(53, \ldots, 59) = \frac{\sum_{t=53}^{59}y_{\text{close}}(t)}{6}$, and the all values of the close price, $\bar{y}_{\text{close}}(1, \ldots, 59) = \frac{\sum_{t=1}^{59}y_{\text{close}}(t)}{59}$, and computed the estimated percentage of change using $\frac{\bar{y}_{\text{close}}(53, \ldots, 59)}{\bar{y}_{\text{close}}(1, \ldots, 59)} - 1$.

To compare the effect of using these proposed features to make trading decisions, we followed the recommendations of \cite{parente2024} and used RSI (Relative Strength Index), ULTOSC (Ultimate Oscillator), Close price percentage change, Z-Score of the close price in a given number of time frames, Z-Score of the volume and the MA (Moving Average) ratio. These features were computed for the entire time series as proposed by \cite{parente2024}, and their values at time $t = 59$ were used to predict $t = 60$, respecting the temporal component and not including any future information on feature calculation.

\subsection*{Clustering markets using Gaussian mixture models}
All Bitcoin, Pepecoin and Nasdaq markets were filtered using a Gaussian Mixture Model (GMM). We selected the number of GMM components (i.e. clusters) based on the BIC criterion, selecting the model with the lowest values of BIC for each market. The parameters of the GMM were estimated using the standardised proposed features and commonly used ones to evaluate possible changes in performance and cluster estimation for both cases. 
\begin{figure}[!ht]
    \centering
    \includegraphics[width=1\linewidth]{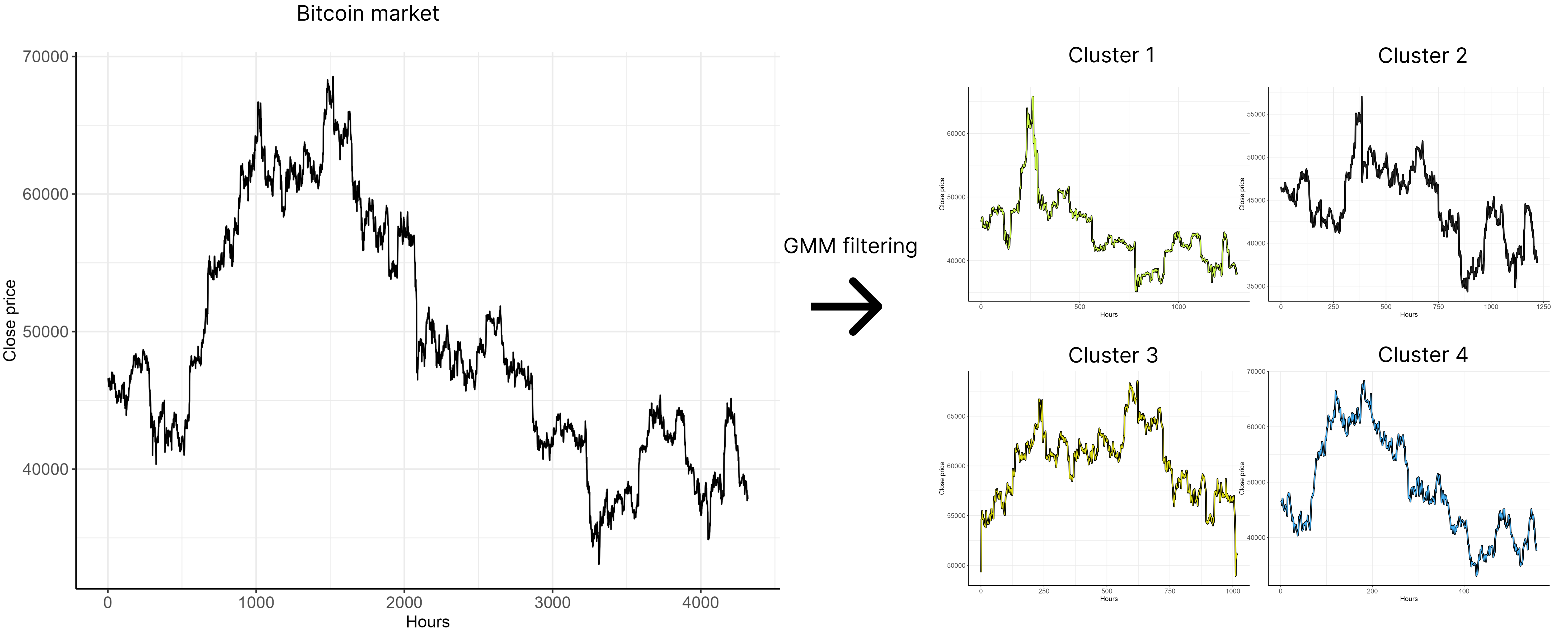}
    \caption{Gaussian mixture model filtering approach applied to the Bitcoin market using the proposed features. The diagram shows the obtained time series clustered into $4$ groups. }
    \label{GMMFiltering}
\end{figure}
The GMM-filtering took into account all the features used for classifying trading decisions. Figure~\ref{GMMFiltering} presents a diagram illustrating this procedure based on the Bitcoin market, which uses the proposed features with $4$ clustered time series selected based on the BIC criterion. The Gaussian mixture model serves as a clustering method that combines similar trends of the financial time series. The hourly observations are rearranged and filtered based on the features collected from the original observations of the studied market. This approach helps test the hypothesis that training machine learning algorithms based on similar market trends can enhance their performance.

\subsection*{Modelling market movements using machine learning}
Before employing any learning algorithm, we defined a baseline algorithm representing a naive prediction where a decision $d_{m}(t)$ for a market $m$ on minute $t$ is obtained, assuming that the three possible decisions are equally likely to be sampled. The naive approach does not consider any features extracted by the markets. This benchmark allows us to evaluate the performance of the various learning algorithms. We used K-nearest neighbours (KNN), Deep Neural Networks (DNN), Polynomial Support Vector Machines (Poly SVM), Random Forests and XGBoost algorithms to classify trading decisions (buy, sell and hold). 

To estimate the performance of these algorithms we computed the number of times the learning algorithm accurately selected a buying ($T_{\text{b}}$), selling ($T_{\text{s}}$) and holding ($T_{\text{h}}$) decision. We also calculated the number of times the learning algorithm misclassified buying as a selling ($F_{\text{bs}}$), buying as a holding ($F_{\text{bh}}$), selling as a buying ($F_{\text{sb}}$) selling as a holding ($F_{\text{sh}}$), holding as a buying ($F_{\text{hb}}$), and holding as a selling ($F_{\text{hs}}$) decision. Based on these metrics we calculated the accuracy for buying, $\text{acc}_{b} = \frac{T_{\text{b}}}{F_{\text{bs}} + T_{\text{b}} + F_{\text{bh}}}$, selling, $\mbox{acc}_{s} = \frac{T_{\text{s}}}{F_{\text{sb}} + T_{\text{s}} + F_{\text{sh}}}$, and holding, $\mbox{acc}_{h} = \frac{T_{\text{h}}}{F_{\text{hb}} + T_{\text{h}} + F_{\text{hs}}}$, categories. Finally, we reported the average of individual accuracies by computing $\bar{\mbox{acc}} = \frac{\text{acc}_{b} + \text{acc}_{s} + \text{acc}_{h}}{3}$.

We also computed a profit performance based on the time series of percentage of change, $\Omega_m$, and selected decisions $\mathbf{D}_m$ for each market $m$ using the function
$$
g(\omega_m(t+1), d_m(t)) = 
\begin{cases}
\omega_m(t) - \omega_m(t+1), & \text{if } d_m(t) = \mbox{Sell} \\
\omega_m(t) + \omega_m(t+1), & \text{if } d_m(t) = \mbox{Buy} \\
\omega_m(t), & \text{otherwise}
\end{cases}$$
where $\omega_m(t)$ is the percentage change and $d_m(t)$ is the predicted decision by a learning algorithm at time $t$. Based on this function we compute the Accumulated Percentage Change $APC = \sum_{t = 1}^{T-1}g(\omega_m(t+1), d_m(t))$. We produced random trading decisions as the benchmark by following the naive algorithm that assigns equal probability to all possible decisions, thus allowing us to compute $\bar{\mbox{acc}}$ and $APC$. We produced $10,000$ values of both performance metrics. A temporal cross-validation approach was applied using the initial percentage of observations to train the learning algorithms. The unseen percentage of the data to obtain $\bar{\mbox{acc}}$ and $APC$ for the algorithms and to obtain $10,000$ values of both performance metrics based on the naive algorithm. We used a temporal train and test approach to obtain the performance of the learning algorithms. The test sets that were evaluated included $20\%$, $30\%$ and $40\%$ of the data. Finally, the Python and R programming languages were used to implement the methods proposed in this paper. To allow full reproducibility of the findings, we have made the code available at \url{https://github.com/GabrielRPalma/ClarrifyingTraderDecisionsWithML}.

\section{Results and discussion}

\subsection*{Gaussian Mixture Model clustering}
Based on the BIC criterion for selecting the number of clusters using the Gaussian Mixture Model approach, the indicated number of clusters when using the proposed features was $4$. On the other hand, when using the commonly used features (RSI, ULTOSC, Close price percentage change, Z-Score of the close price, volume Z-Score, and Moving average ratio), a single unique cluster was indicated based on the BIC values. As a result, we couldn't use the proposed filtering strategy when using these features, considering that for each analysed market, all observations belong to a single cluster. The obtained clusters in both cases had BIC values approximately $30\%$ smaller than the second best number of clusters

\begin{table}[ht]
\caption{Gaussian Mixture Model (GMM) clustering averages based on the standardised commonly used features collected in 59-minute intervals for Bitcoin, Pepecoin and Nasdaq markets.}
\centering
\begin{tabular}{lrrr}
\hline
Features & Bitcoin & Pepecoin & Nasdaq \\
\hline
Close price & 0.45 & 0.15 & 0.57 \\ 
Volume & 0.01 & 0.02 & 0.05 \\ 
Close price Z-score & 0.52 & 0.49 & 0.50 \\ 
RSI & 0.48 & 0.01 & 0.49 \\ 
ULTOSC & 0.47 & 0.34 & 0.49 \\ 
Percentage of change & 0.67 & 0.30 & 0.36 \\ 
Volume Z-score& 0.01 & 0.02 & 0.05 \\ 
Moving average ratio & 0.44 & 0.36 & 0.50 \\ 

\hline
\label{GMMAveragesTableCommon}
\end{tabular}
\end{table}

Table~\ref{GMMAveragesTableCommon} presents the standardised mean values of the commonly used features collected at 59-minute intervals for the single estimated cluster derived using the Gaussian Mixture Model (GMM) method. The Nasdaq market cluster is characterized by higher price levels, the Bitcoin cluster by higher price movements, and the Pepecoin cluster by lower RSI, ULTOSC, and moving average ratio values. This result highlights the importance of the feature engineering process when exploring the capabilities of learning algorithms in different market scenarios based on GMM filtering.

Table~\ref{GMMAveragesTable} presents the standardised mean values of the proposed features collected using 59-minute intervals for the four estimated clusters derived using the Gaussian Mixture Model (GMM) method. For the Bitcoin market, the clearer differences are in the close price, linear model intercept, and peaks average magnitude features. Cluster $3$ has substantially higher mean values for these three features ($0.77$, $0.76$, $0.76$ respectively) compared to the other clusters, suggesting that this cluster represents time periods with higher price levels. In contrast, clusters $1$, $2$ and $4$ have lower and relatively similar mean values for these price-related features. Also, cluster $4$ has the highest proportion of buying and selling behaviour, indicating more decisions based on the proposed threshold algorithm. 

For the Pepecoin cryptocurrency, our results showed that the most notable differences are observed in the buy and sell proportions, with cluster $3$ having the highest mean values ($0.67$ and $0.66$, respectively), followed by cluster $1$ ($0.63$ and $0.62$). Clusters 2 and 4 have lower mean values for these features, with cluster 2 being the lowest ($0.41$ and $0.43$). Cluster 3 has the highest mean values for the close price, linear model intercept, and peak average magnitude features ($0.31$, $0.34$, and $0.33$, respectively), while cluster 2 has the lowest (0.06 for all three features). Overall, these differences among clusters for the Pepecoin market are driven mainly by close price values, which is more evident than for the Bitcoin Market.

The more explicit differences for Nasdaq clusters are the buy and sell proportions, which increase from cluster $1$ to cluster $4$, and the price-related features (close price, linear model intercept, and peaks average magnitude), which decrease from cluster $1$ to cluster $4$. Our findings suggest an inverse relationship between trading activity and price levels among estimated Nasdaq clusters.

\begin{table}[ht]
\caption{Gaussian Mixture Model (GMM) clustering averages based on the standardised proposed features collected in 59-minute intervals for Bitcoin, Pepecoin and Nasdaq markets.}
\centering
\begin{tabular}{llrrrr}
\hline
Markets & Features & Cluster 1 & Cluster 2 & Cluster 3 & Cluster 4 \\
\hline
&Buy proportion & 0.43 & 0.55 & 0.52 & 0.60 \\
&Sell proportion & 0.47 & 0.58 & 0.52 & 0.60 \\
&Close price & 0.33 & 0.33 & 0.77 & 0.45 \\
Bitcoin &Linear model intercept & 0.32 & 0.32 & 0.76 & 0.45 \\
&Linear model slope & 0.55 & 0.56 & 0.55 & 0.57 \\
&Peaks average curvature & 0.95 & 0.90 & 0.89 & 0.84 \\
&Peaks average magnitude & 0.32 & 0.32 & 0.76 & 0.45 \\
&Estimated percentage change & 0.43 & 0.44 & 0.43 & 0.43 \\[0.15cm]

&Buy proportion & $0.63$ & $0.41$ & $0.67$ & $0.53$ \\
&Sell proportion & $0.62$ & $0.43$ & $0.66$ & $0.54$ \\
&Close price & $0.22$ & $0.06$ & $0.31$ & $0.17$ \\
&Linear model intercept & $0.24$ & $0.06$ & $0.34$ & $0.18$ \\
Pepecoin&Linear model slope & $0.31$ & $0.31$ & $0.34$ & $0.32$ \\
&Peaks average curvature & $0.96$ & $0.99$ & $0.86$ & $0.98$ \\
&Peaks average magnitude & $0.24$ & $0.06$ & $0.33$ & $0.18$ \\
&Estimated percentage change & $0.29$ & $0.29$ & $0.30$ & $0.29$ \\[0.15cm]

&Buy proportion & $0.40$ & $0.54$ & $0.63$ & $0.69$ \\
&Sell proportion & $0.41$ & $0.53$ & $0.61$ & $0.63$ \\
&Close price & $0.68$ & $0.59$ & $0.50$ & $0.43$ \\
&Linear model intercept & $0.68$ & $0.59$ & $0.50$ & $0.43$ \\
Nasdaq&Linear model slope & $0.45$ & $0.46$ & $0.47$ & $0.47$ \\
&Peaks average curvature & $0.96$ & $0.92$ & $0.84$ & $0.70$ \\
&Peaks average magnitude & $0.68$ & $0.59$ & $0.50$ & $0.43$ \\
&Estimated percentage change & $0.59$ & $0.59$ & $0.60$ & $0.58$ \\
\hline
\label{GMMAveragesTable}
\end{tabular}
\end{table}

\subsection*{Trading decision classification with ML}
\begin{figure}
    \centering
    \includegraphics[width=1\linewidth]{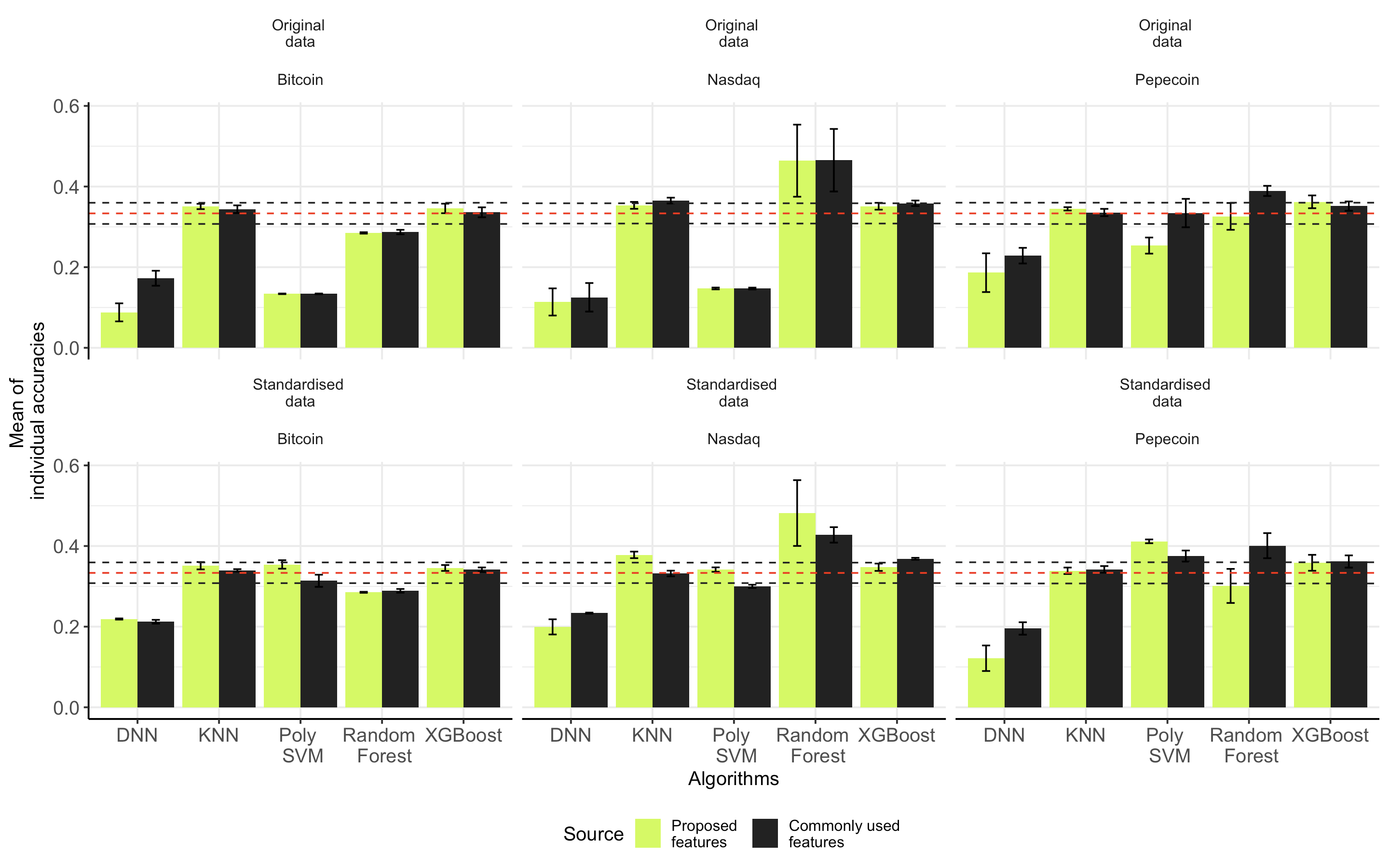}
    \caption{The mean of individual accuracies of the ML algorithms per studied market, when using the original and standardised data. The red and black dashed lines represent the average, the $2.5\%$ and $97.5\%$ percentiles of the performance metric using the random algorithm.}
    \label{IndividualAcuracies}
\end{figure}

Figure~\ref{IndividualAcuracies} shows the performance of the learning algorithms in classifying trading decisions for all markets. The figure also shows the performance metric variation based on the different testing percentages used ($20\%$, $30\%$, and $40\%$) as shown by the error bar computed using the standard error of these metrics for all testing percentages. Overall, the only algorithms that produced higher performance than the $97.5\%$ percentile random algorithm threshold were the Random Forests and KNN algorithms. The DNN and polynomial SVM algorithms produced higher average performances when using the standardised data in most scenarios.

None of the learning algorithms achieved higher performance metrics for the Bitcoin market than the $97.5\%$ percentile of the random algorithm. However, XGBoost and KNN consistently achieved higher average performance. Also, the polynomial SVM with its parameters estimated using the standardised data obtained higher performance when using the proposed features than the commonly used ones. Furthermore, the Random Forests and KNN obtained higher average performances than the $97.5\%$ percentile of the random algorithm for the Nasdaq market. Our results also showed that the Random Forests algorithm trained with the standardised proposed features obtained higher performance than the commonly used features. Finally, the polynomial SVM and Random Forests also obtained higher performances when compared to the random algorithm for the Pepecoin market. 

\begin{figure}
    \centering
    \includegraphics[width=1\linewidth]{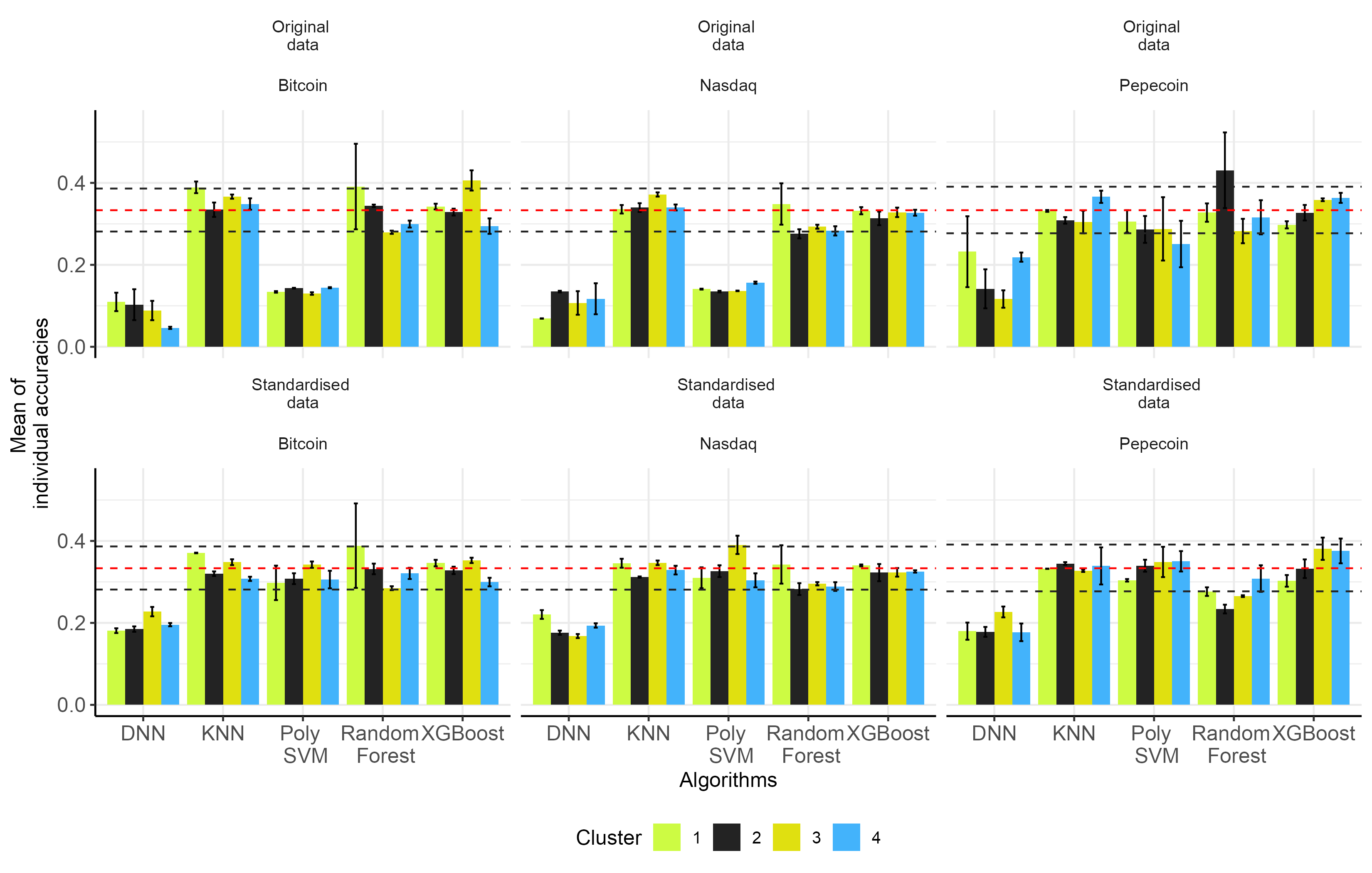}
    \caption{The mean of individual accuracies of the ML algorithms per GMM filtered time series based on the GMM clusters estimated with the proposed features, when using the original and standardised data. The red and black dashed lines represent the average, the $2.5\%$ and $97.5\%$ percentiles of the performance metric using the random algorithm.}
    \label{ClusteredIndividualAcuracies}
\end{figure}

Figure~\ref{ClusteredIndividualAcuracies} shows the average of individual accuracies of the learning algorithms for the GMM-filtered markets based on the proposed features. Overall, the KNN, polynomial SVM, Random Forests and XGBoost algorithms achieved higher average performance than the $97.5\%$ percentile of the random algorithm. Also, the DNN and polynomial SVM algorithms increased their average performance when training with the standardised data for most clusters and markets. Our findings showed no apparent increase in the average performance after applying the filtering strategy based on the GMM. However, the number of algorithms that produced higher performance than the random algorithm increased after applying the filtering technique. The Bitcoin market time series filtered with the first and third clusters achieved higher performance according to the average of individual accuracies when using the Random Forests and XGBoost algorithms. The polynomial SVM algorithm only achieved higher performance using the Nasdaq market time series filtered with the third cluster. Finally, the Random Forests algorithm produced higher performance on the Pepecoin market time series filtered with the second cluster. 

\begin{figure}
    \centering
    \includegraphics[width=1\linewidth]{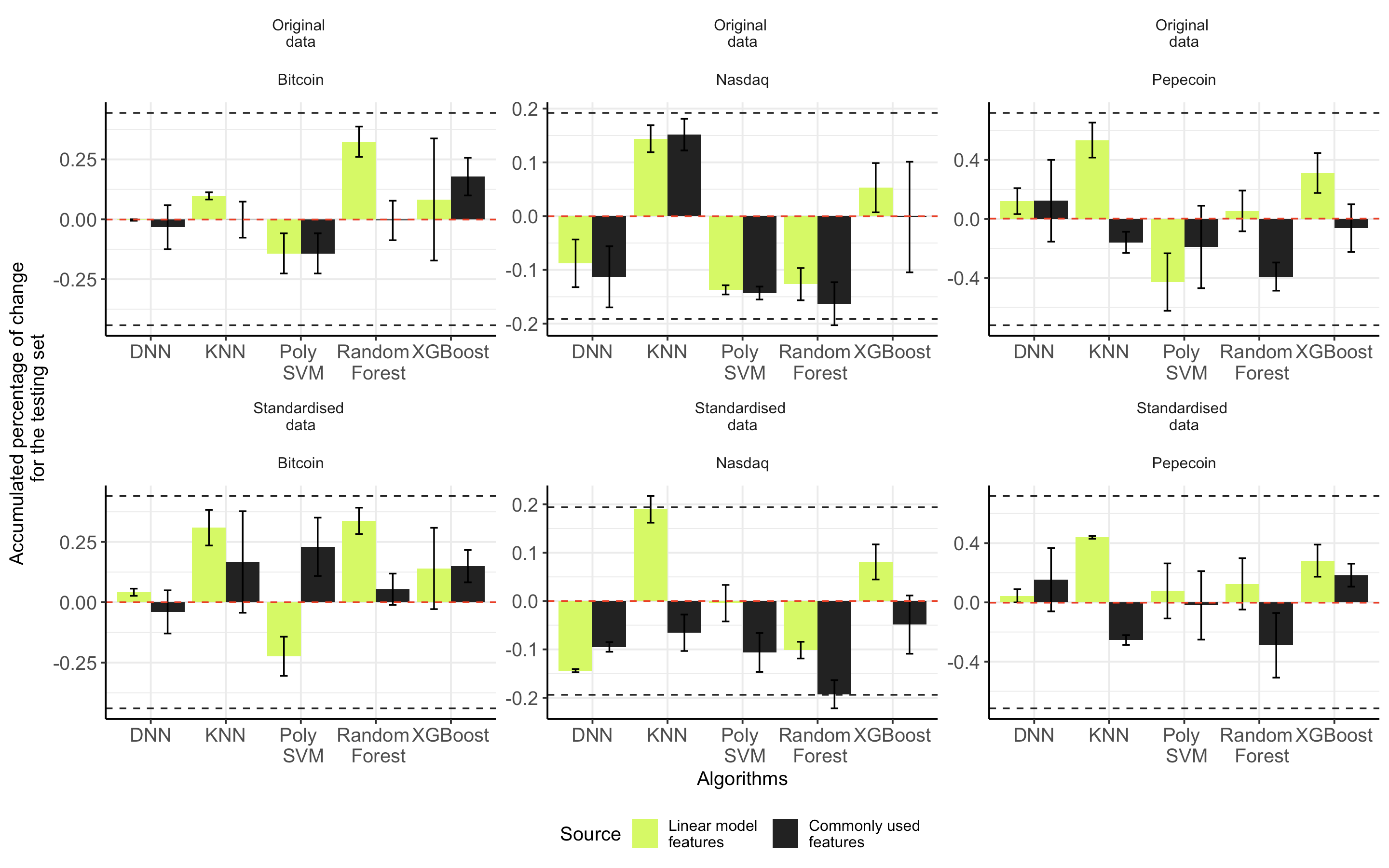}
    \caption{The Accumulated percentage of change of the ML algorithms per studied market, when using the original and standardised data. The red and black dashed lines represent the average, the $2.5\%$ and $97.5\%$ percentiles of the performance metric using the random algorithm.}
    \label{AccumulatedChange}
\end{figure}

Figure~\ref{AccumulatedChange} shows the Accumulated Percentage of Change (APC) when using the classified decision by the learning algorithms using $20\%$, $30\%$, and $40\%$ of the observations for testing. Overall, none of the algorithms produced a higher average performance than the $97.5\%$ percentile of the metrics obtained based on the random algorithm for all markets. The KNN algorithm only achieved the closest value to this threshold when including the variation based on the different testing sets. Our results highlight a benefit of using the standardised data for the Bitcoin market, which increases the APC in most of the algorithms. The highest APC obtained was with the Random Forests algorithm based on the proposed features. On the other hand, the standardisation of the features did not increase the overall APC of the studied algorithms for the Nasdaq market. The highest APC obtained was with the KNN algorithm, which also used the proposed features. The KNN algorithm trained with the proposed features produced the highest APC for the Pepecoin market, and we did not find any clear improvement when standardising the features.

\begin{figure}
    \centering
    \includegraphics[width=1\linewidth]{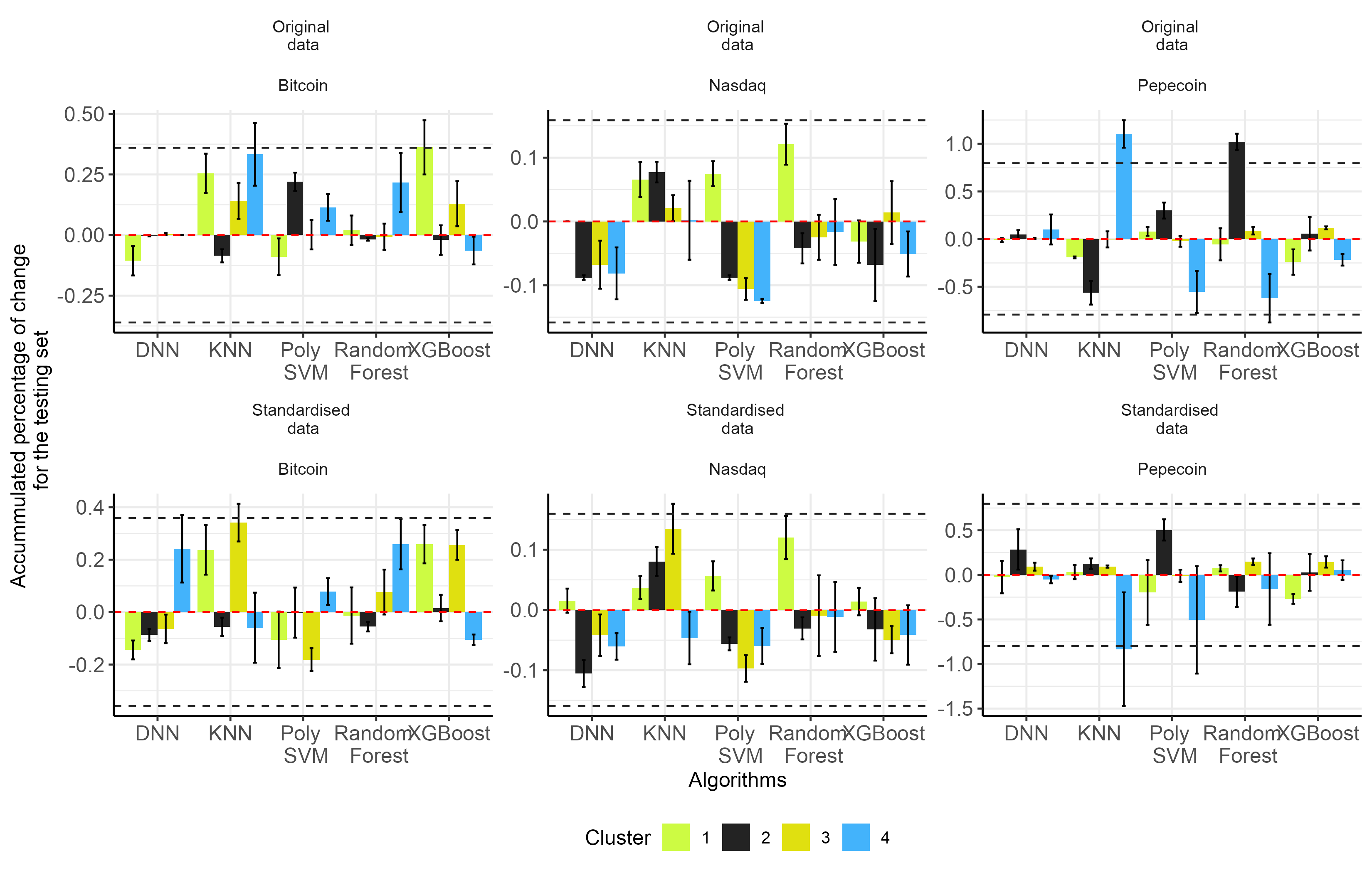}
    \caption{The Accumulated percentage of change of the ML algorithms per GMM filtered time series based on the GMM clusters estimated with the proposed features, when using the original and standardised data. The red and black dashed lines represent the average, the $2.5\%$ and $97.5\%$ percentiles of the performance metric using the random algorithm.}
    \label{ClusteredAccumulatedChange}
\end{figure}

Figure~\ref{ClusteredAccumulatedChange} shows the APC of the learning algorithms for the GMM-filtered markets based on the proposed features. Overall, the number of algorithms that obtained higher APC values than the $97.5\%$ percentile of the random algorithm increased, with the KNN, Random Forests and XGBoost highlighting this finding. The highest APC was obtained by the KNN algorithm using the original proposed features filtered by cluster $4$. The XGBoost algorithm produced the better average APC based on the original proposed features filtered by cluster 1 for the Bitcoin market, while the KNN algorithm produced higher APC values based on the standardised proposed features filtered by cluster $3$ for the Nasdaq market. Finally, the KNN algorithm produced higher APC values based on the original proposed features filtered by cluster $4$ for the Pepecoin market.

Our results highlight the competitive nature of the proposed features compared to ones commonly used to predict market movements, with APC values higher than the random algorithm. The potential of the GMM-filtering approach is highlighted by the finding that the number of algorithms which produced better performance in terms of mean individual accuracies and APC values increased after the application of the procedure. The reported algorithms that produced these superior performances (KNN, Random Forests and XGBoost) were already mentioned in previous work showing promising results \citep{GERLEIN2016, hasan2020, prasad2021, JAQUART202145, parente2024, BOUTESKA2024}.

The use of the Gaussian mixture model is a common practice in financial data analysis \citep{eirola2013gaussian, bai2014clustering}, and this method has already been used as a pre-processing step for improving classification in different research areas \citep{fan2022preprocessing}. Other techniques have also been proposed as a pre-processing step, such as the principal component analysis and association metrics based clustering \citep{palma2023pattern, fan2024using}.  Our results illustrate the feasibility of applying GMM filtering to make trading decisions \citep{buczynski2021, SOKOLOVSKY2023}. This work can serve as the basis for the future development of feature engineering tools with application to financial markets \citep{long2019deep, dami2021predicting, cerqueira2021vest, mantilla2023novel, verdonck2021special, zhang2024deep}.

\section{Conclusion}
We explored the capabilities of classic machine learning methods to make trading decisions in emergent and well-established markets using a new set of features based on a linear model of close prices and their curvatures. We also proposed applying Gaussian mixture models to filter these markets to better explore the capabilities of these learning algorithms in finding profit opportunities. Our results revealed the potential of the proposed features, showing that filtering the time series enhances average performance over a naive algorithm that assigns equal probability for each trading decision. When applying the proposed filtering technique with GMM to the Pepecoin market, we found that the KNN and Random Forests algorithms produced significantly higher average returns than the random algorithm.

\section{Acknowledgments}

This publication has resulted from research conducted with the financial support of DLT Capital and Science Foundation Ireland under Grant 18/CRT/6049. The opinions, findings and conclusions or recommendations expressed in this material are those of the authors and do not necessarily reflect the views of the funding agencies.

\section{Declarations}
~~~~
\textbf{Ethical Approval} Not applicable.

\textbf{Competing interests} Not applicable.

\textbf{Authors’ contributions} G.R.P. conceived and designed the research. P.M. and M.S. collected the data and provided insights into the discussion of results. G.R.P. created the approach and analysed the data. G.R.P. led the writing of the manuscript. All authors contributed to the overall writing. 

\textbf{Funding} Science Foundation Ireland under Grant 18/CRT/6049 and DLT Capital. 


\bibliographystyle{apalike}
\bibliography{ref}

\begin{thebibliography}{}

\bibitem[Avramelou et~al., 2024]{avramelou2024deep}
Avramelou, L., Nousi, P., Passalis, N., and Tefas, A. (2024).
\newblock Deep reinforcement learning for financial trading using multi-modal features.
\newblock {\em Expert Systems with Applications}, 238:121849.

\bibitem[Bai and Dang, 2014]{bai2014clustering}
Bai, Y. and Dang, J. (2014).
\newblock Clustering analysis of stock volume and price relationship based on gaussian mixture model.
\newblock In {\em 2014 International Conference on Mechatronics, Electronic, Industrial and Control Engineering (MEIC-14)}, pages 1552--1555. Atlantis Press.

\bibitem[Bouteska et~al., 2024]{BOUTESKA2024}
Bouteska, A., Abedin, M.~Z., Hajek, P., and Yuan, K. (2024).
\newblock Cryptocurrency price forecasting – a comparative analysis of ensemble learning and deep learning methods.
\newblock {\em International Review of Financial Analysis}, 92:103055.

\bibitem[Buczynski et~al., 2021]{buczynski2021}
Buczynski, W., Cuzzolin, F., and Sahakian, B. (2021).
\newblock A review of machine learning experiments in equity investment decision-making: why most published research findings do not live up to their promise in real life.
\newblock {\em International Journal of Data Science and Analytics}, 11:221--242.

\bibitem[Cerqueira et~al., 2021]{cerqueira2021vest}
Cerqueira, V., Moniz, N., and Soares, C. (2021).
\newblock Vest: Automatic feature engineering for forecasting.
\newblock {\em Machine Learning}, pages 1--23.

\bibitem[Cheng and Chen, 2023]{cheng2023general}
Cheng, T. and Chen, K. (2023).
\newblock A general framework for portfolio construction based on generative models of asset returns.
\newblock {\em The Journal of Finance and Data Science}, 9:100113.

\bibitem[Dami and Esterabi, 2021]{dami2021predicting}
Dami, S. and Esterabi, M. (2021).
\newblock Predicting stock returns of tehran exchange using lstm neural network and feature engineering technique.
\newblock {\em Multimedia Tools and Applications}, 80(13):19947--19970.

\bibitem[Dash and Dash, 2016]{dash2016hybrid}
Dash, R. and Dash, P.~K. (2016).
\newblock A hybrid stock trading framework integrating technical analysis with machine learning techniques.
\newblock {\em The Journal of Finance and Data Science}, 2(1):42--57.

\bibitem[De~Luca and Zuccolotto, 2014]{de2014dynamic}
De~Luca, G. and Zuccolotto, P. (2014).
\newblock Dynamic clustering of financial assets.
\newblock In {\em Analysis and Modeling of Complex Data in Behavioral and Social Sciences}, pages 103--111. Springer.

\bibitem[Eirola and Lendasse, 2013]{eirola2013gaussian}
Eirola, E. and Lendasse, A. (2013).
\newblock Gaussian mixture models for time series modelling, forecasting, and interpolation.
\newblock In {\em Advances in Intelligent Data Analysis XII: 12th International Symposium, IDA 2013, London, UK, October 17-19, 2013. Proceedings 12}, pages 162--173. Springer.

\bibitem[Fan et~al., 2022]{fan2022preprocessing}
Fan, C., Zhang, N., Jiang, B., and Liu, W.~V. (2022).
\newblock Preprocessing large datasets using gaussian mixture modelling to improve prediction accuracy of truck productivity at mine sites.
\newblock {\em Archives of Mining Sciences}, pages 661--680.

\bibitem[Fan et~al., 2024]{fan2024using}
Fan, C., Zhang, N., Jiang, B., and Liu, W.~V. (2024).
\newblock Using deep neural networks coupled with principal component analysis for ore production forecasting at open-pit mines.
\newblock {\em Journal of Rock Mechanics and Geotechnical Engineering}, 16(3):727--740.

\bibitem[Fang et~al., 2022]{fang2022cryptocurrency}
Fang, F., Ventre, C., Basios, M., Kanthan, L., Martinez-Rego, D., Wu, F., and Li, L. (2022).
\newblock Cryptocurrency trading: a comprehensive survey.
\newblock {\em Financial Innovation}, 8(1):13.

\bibitem[Gerlein et~al., 2016]{GERLEIN2016}
Gerlein, E.~A., McGinnity, M., Belatreche, A., and Coleman, S. (2016).
\newblock Evaluating machine learning classification for financial trading: An empirical approach.
\newblock {\em Expert Systems with Applications}, 54:193--207.

\bibitem[Granha et~al., 2023]{granha2023three}
Granha, M.~F., Zubillaga~Herrera, B., Vilela, A.~L., Wang, C., and Stanley, H. (2023).
\newblock Three-state opinion dynamics for financial markets on complex networks.
\newblock In {\em APS March Meeting Abstracts}, volume 2023, pages D47--011.

\bibitem[Hasan et~al., 2020]{hasan2020}
Hasan, A., Kal{\i}ps{\i}z, O., and Akyoku{\c{s}}, S. (2020).
\newblock Modeling traders’ behavior with deep learning and machine learning methods: evidence from bist 100 index.
\newblock {\em Complexity}, 2020(1):8285149.

\bibitem[Jaquart et~al., 2021]{JAQUART202145}
Jaquart, P., Dann, D., and Weinhardt, C. (2021).
\newblock Short-term bitcoin market prediction via machine learning.
\newblock {\em The Journal of Finance and Data Science}, 7:45--66.

\bibitem[Kwon and Lee, 2024]{kwon2024hybrid}
Kwon, Y. and Lee, Z. (2024).
\newblock A hybrid decision support system for adaptive trading strategies: Combining a rule-based expert system with a deep reinforcement learning strategy.
\newblock {\em Decision Support Systems}, 177:114100.

\bibitem[Lin et~al., 2021]{lin2021improving}
Lin, Y., Liu, S., Yang, H., Wu, H., and Jiang, B. (2021).
\newblock Improving stock trading decisions based on pattern recognition using machine learning technology.
\newblock {\em PloS one}, 16(8):e0255558.

\bibitem[Long et~al., 2019]{long2019deep}
Long, W., Lu, Z., and Cui, L. (2019).
\newblock Deep learning-based feature engineering for stock price movement prediction.
\newblock {\em Knowledge-Based Systems}, 164:163--173.

\bibitem[Mantilla and Dormido-Canto, 2023]{mantilla2023novel}
Mantilla, P. and Dormido-Canto, S. (2023).
\newblock A novel feature engineering approach for high-frequency financial data.
\newblock {\em Engineering Applications of Artificial Intelligence}, 125:106705.

\bibitem[Paiva et~al., 2019]{paiva2019decision}
Paiva, F.~D., Cardoso, R. T.~N., Hanaoka, G.~P., and Duarte, W.~M. (2019).
\newblock Decision-making for financial trading: A fusion approach of machine learning and portfolio selection.
\newblock {\em Expert Systems with Applications}, 115:635--655.

\bibitem[Palma et~al., 2023a]{palma2023pattern}
Palma, G.~R., Godoy, W.~A., Engel, E., Lau, D., Galvan, E., Mason, O., Markham, C., and Moral, R.~A. (2023a).
\newblock Pattern-based prediction of population outbreaks.
\newblock {\em Ecological Informatics}, 77:102220.

\bibitem[Palma et~al., 2023b]{palma2023}
Palma, G.~R., Thornberry, C., Commins, S., and Moral, R. d.~A. (2023b).
\newblock Understanding learning from eeg data: Combining machine learning and feature engineering based on hidden markov models and mixed models.
\newblock {\em arXiv preprint arXiv:2311.08113}.

\bibitem[Parente et~al., 2024]{parente2024}
Parente, M., Rizzuti, L., and Trerotola, M. (2024).
\newblock A profitable trading algorithm for cryptocurrencies using a neural network model.
\newblock {\em Expert Systems with Applications}, 238:121806.

\bibitem[Prasad and Seetharaman, 2021]{prasad2021}
Prasad, A. and Seetharaman, A. (2021).
\newblock Importance of machine learning in making investment decision in stock market.
\newblock {\em Vikalpa}, 46(4):209--222.

\bibitem[Sebasti{\~a}o and Godinho, 2021]{sebastiao2021forecasting}
Sebasti{\~a}o, H. and Godinho, P. (2021).
\newblock Forecasting and trading cryptocurrencies with machine learning under changing market conditions.
\newblock {\em Financial Innovation}, 7:1--30.

\bibitem[Singh et~al., 2024a]{singh2024time}
Singh, H., Sharma, C., Attri, V., and Singh, S. (2024a).
\newblock Time series forecast with stock's price candlestick patterns and sequence similarities.
\newblock In {\em 2024 International Conference on Emerging Smart Computing and Informatics (ESCI)}, pages 1--6. IEEE.

\bibitem[Singh et~al., 2024b]{singh2024ensemble}
Singh, H., Sharma, P., Prabha, C., Singh, S., et~al. (2024b).
\newblock Ensemble learning with an adversarial hypergraph model and a convolutional neural network to forecast stock price variations.
\newblock {\em Ing{\'e}nierie des Syst{\`e}mes d'Information}, 29(3).

\bibitem[Sokolovsky et~al., 2023]{SOKOLOVSKY2023}
Sokolovsky, A., Arnaboldi, L., Bacardit, J., and Gross, T. (2023).
\newblock Interpretable trading pattern designed for machine learning applications.
\newblock {\em Machine Learning with Applications}, 11:100448.

\bibitem[Verdonck et~al., 2021]{verdonck2021special}
Verdonck, T., Baesens, B., {\'O}skarsd{\'o}ttir, M., and vanden Broucke, S. (2021).
\newblock Special issue on feature engineering editorial.
\newblock {\em Machine learning}, pages 1--12.

\bibitem[Zhang et~al., 2024]{zhang2024deep}
Zhang, C., Sjarif, N. N.~A., and Ibrahim, R. (2024).
\newblock Deep learning models for price forecasting of financial time series: A review of recent advancements: 2020--2022.
\newblock {\em Wiley Interdisciplinary Reviews: Data Mining and Knowledge Discovery}, 14(1):e1519.

\bibitem[Zou et~al., 2024]{zou2024novel}
Zou, J., Lou, J., Wang, B., and Liu, S. (2024).
\newblock A novel deep reinforcement learning based automated stock trading system using cascaded lstm networks.
\newblock {\em Expert Systems with Applications}, 242:122801.

\end{thebibliography}

\end{document}